\documentstyle[twocolumn,aps]{revtex}

\def\fun#1#2{\lower3.6pt\vbox{\baselineskip0pt\lineskip.9pt
\ialign{$\mathsurround=0pt#1\hfil##\hfil$\crcr#2\crcr\sim\crcr}}}

\begin{document}

\title{Model independent radiative corrections in processes  of
polarized\\
electron-nucleon elastic scattering}

\author{A.~Afanasev$^{a}$,I.~Akushevich$^{a}$, N.~Merenkov$^{b}$\\
a) North Carolina Central University,  
Durham, NC 27707 and
TJNAF, Newport News, VA 23606\\
b) Kharkov Institute of Physics and Technology, Kharkov 310108, Ukraine
}

\date{\today}

\input{epsf}

%\begin{center} {\Large \bf Status: Very preliminary} \end{center}

\maketitle

\begin{abstract}
The explicit formulae for radiative correction (RC) calculation for
elastic
$ep$-scattering is presented. Two typical measurements of polarization
observables such as beam-target asymmetry or recoil proton polarization,  
are considered.
Possibilities to take into account realistic experimental acceptances are
discussed. The Fortran code MASCARAD for providing the RC procedure is
presented. Numerical analysis is done for kinematical conditions of
TJNAF.
\end{abstract}
\pacs{Valid PACS appear here.
{\tt$\backslash$\string pacs\{\}} should always be input,
even if empty.}

\narrowtext

\section{Introduction}

%CEBAF and experiments.

Precise polarization measurements of nucleon form factors in electron
scattering
is an essential component of new-generation electron accelerators such as
CEBAF.
This unprecedented precision requires knowledge of higher-order 
electromagnetic effects at a per-cent level. The purpose of our work is to
analyze
radiative corrections in elastic electron proton scattering and develop
proper 
computational techniques that could be used  in experiments at Jefferson
and
other electron
accelerator labs.

The %abnormally
large statistics obtained in the experiments at CEBAF \cite{1CEBAFothers}
requires adequate taking into account radiative effects.

Both spin averaged and spin dependent parts of the cross section
($\sigma^u$ and $\sigma^p$) can be
presented as
\begin{equation}\label{first}
\sigma_{obs}^{u,p}=(1+\delta)\sigma_0^{u,p}+\sigma_R^{u,p}.
\end{equation}
Both the factorized correction $\delta$ and unfactorized cross section coming
from bremsstrahlung process contribute to cross section.
When polarization asymmetries of elastic processes are considered,
the factorized part of total RC has tendency to cancel but
unfactorized part can give important contribution. This is the
reason why we pay special attention to the effects which have
different contributions to polarized and unpolarized parts of
elastic cross section, namely
\begin{itemize}
\item A contribution of hard radiation. Its contribution is
restricted by using of kinematical cuts on deviation from
energy-momentum conservation.
\item A higher order correction, which contribute to the polarization
asymmetry (due to unfactorized effects) even at the leading order.
\item A contribution of box-type diagrams which are a typical
example of the unfactorized correction.
\end{itemize}

First and second correction are so-called model independent RC.	They do
not depend on any assumptions on nuclear interactions. In our case it
means that model independent correction can be written in terms of
elastic nucleon formfactors. 
%The third correction requires the
%consideration of two-boson exchange graph, which can be calculated
%without any additional assumptions.
The third correction is model dependent because it requires the
consideration of two--photon exchange graphs, which cannot be
calculated without any additional assumptions. Its calculation in the
soft
photon approximation,
which gives identical corrections to both polarized and unpolarized parts
of the cross sections, was done in refs.\cite{Maximon1,Maximon2}.  

A modern approach of solving the task on RC calculation assumes the exact
calculation of the lowest order model independent correction. The task
can be solved exactly. Uncertainties of
the model independent RC can come only from fits and models used for
structure functions. The second order model independent correction also
can be performed exactly, however usually it leads to cumbersome or lengthy
 formulae, and
realistic approximations are preferred. In this case there is
an additional uncertainty coming from using approximate formulae instead of
exact ones. The calculation of model dependent	correction requires
additional assumptions about hadron interaction, so it has additional
pure theoretical uncertainties, which are hardly controlled. So
usually only those model dependent correction are considered, which are
expected to strongly influence extracted observable quantities.
In our case such a correction is box-type graphs.

There are two basic methods of calculation of model independent QED
radiative correction. First one is connected with introducing of an
artificial parameter separating the momentum phase space into soft and
hard
parts. One can find a classical review introducing this
formalism in ref.\cite{MoTsai}. However the presence of the artificial
parameter is a disadvantage of this method. For the soft photon part
 the calculation is performed in soft photon
approximation, when its energy is considered to be small with respect all
momenta and masses in the task. So this parameter should be chosen as
small as
possible to reduce the region evaluated approximately. However it cannot
be chosen too small because numerical instabilities of hard part
calculation can occur. In the end of seventies Bardin and Shumeiko
developed an approach \cite{BSh} of extraction and cancellation of
infrared divergence without introducing this artificial
parameter
\footnote{Toll for that is non-positively defined hard
photon contribution, which can lead to difficulties for MC generation
see \cite{RADGEN}. In this paper detailed comparison of explicit formulae
obtained within two considered approaches is given for the case of deep
inelastic scattering.}. Later a lot of calculations were performed within
this approach and a lot of FORTRAN codes created to deal with numerical
calculations. The most known of them are TERAD and POLRAD. The detailed
review of the approach is presented in ref.\cite{Bardin}. In this paper we
use this
approach to calculate the RC of the lowest order to transferred
polarization and asymmetry in elastic electron-proton scattering.
The method allows to calculate the correction exactly. However we will
neglect the lepton mass $m$ everywhere where it is possible.

The observed cross section of the process
\begin{equation}\label{process}
 e(k_1) + N (p) \longrightarrow e'(k_2) + N(p_2),
\end{equation}
is described by one
non-trivial variable, which is usually chosen to be squared of
momentum transferred. There are two ways to reconstruct the variables,
when both lepton and nucleon final momenta are measured. In the first
case it will be denoted as  $Q^2_l=-(k_1-k_2)^2$, and for second case it
is
    $Q^2_h=-(p_2-p)^2$. It is clear that there is no difference
between these definition at the Born level. However at the level of RC
the symmetry is breaking because the radiation of the lepton is
considered only. We consider both cases in this paper.
In the first case the structure of bremsstrahlung cross section looks
like
\begin{equation}
{d\sigma \over d Q^2_l} \sim \alpha^3 \int
{d^3 k \over k_0} \sum {\cal K} {\cal F}^2(Q^2_h) {\cal A}
\label{Q2l}
\end{equation}
where $\cal K$	is a kinematical coefficient calculatable exactly in the
lowest order. It depends on photon variables. ${\cal F}^2$ is a bilinear
combination of nucleon
formfactors dependent on $Q^2_h$ only, which is function of photon
momentum. Usually only final momenta are measured in the part of the
space. It is controlled by function of acceptance ${\cal A}$, 
which is 1 or
0 in depending whether the final particles make it in detectors or not. 
The integral
(\ref{Q2l}) should not be analytically calculated for two reasons.
The first one is dependence of formfactors on $Q^2_h$. We avoid to use some
specific model for them. The second one is acceptance usually very
complicated function of kinematical variable dependent on photon
momentum.

For the second version of reconstruction of transfer momentum squared
the structure of the cross section is
\begin{equation}
{d\sigma \over d Q^2_h} \sim \alpha^3 \sum {\cal F}^2(Q^2_h)\int
{d^3 k \over k_0} {\cal K}  {\cal A}
\label{Q2h}
\end{equation}
In this case the squared formfactor is not dependent on photon momentum and
for 4$\pi$ kinematics (${\cal A}=1$) this integral can be calculated
analytically. In the experimental conditions at JLab \cite{1CEBAFothers},
both
final electron and proton were detected in order to reduce background. However
elastic scattering kinematics was restored by the final proton kinematics, while
electron momentum was integrated over. Therefore, formalism related with
Eq.(\ref{Q2h}) applies for this case.

The calculation of RC to the asymmetry in measurement of recoil
proton for leading and next-to-leading levels
was done within the method of structure functions \cite{AAM}. This method
was chosen because it allows to take into account the higher order
correction simultaneously with enough accuracy of the lowest order RC.
However only the case of measurement of final lepton was considered
there. The traditional calculation based on a covariant method of
infrared divergence cancellation is performed for the two experimental
measurements of polarization observables:
\begin{itemize}
\item measurement of polarization asymmetry within leptonic variables;
\item measurement of asymmetry in recoil proton within hadronic
variables.
\end{itemize}
 This approach allows to
take into account the lowest order RC exactly and to calculate RC within
experimental cuts.
The treatment of the box-type correction and their
influence on transferred polarization requires different methods and
will be a subject of a separate investigation.

\section{Kinematics and Born process}

The Born cross section of the process (\ref{process})
can be written in the form
\begin{equation}
d\sigma_0=\frac{M_0^2}{4pk_1}d\Gamma_0=M_0^2\frac{dQ^2}{16\pi S^2},
\end{equation}
where $S=2k_1p$. Kinematical limits for $Q^2$ are defined as
\begin{equation}
0\leq Q^2 \leq \frac{\lambda_s}{S+m^2+M^2}, \quad \lambda_s=S^2-4m^2M^2,
\end{equation}
where ($m,M$ are the electron and proton masses).
%\begin{equation}
%s=S+m^2+M^2, \qquad \lambda_s=S^2-4m^2M^2
%\end{equation}
    Due to axial symmetry the integration over
azimuthal angle $\phi$ can be performed analytically. However in our
case
kinematical cuts are dependent on this angle so we will consider
two-dimensional Born cross section
\begin{equation}
d\sigma_0=\frac{M_0^2}{4pk_1}d\Gamma_0=M_0^2\frac{dQ^2d\phi}{32\pi^2
S^2}
\label{phiall}
\end{equation}

Born matrix element is
\begin{equation}
M^2=\frac{e^4}{Q^4}L^0_{\mu\nu}W_{\mu\nu}
\end{equation}
We use standard definitions for (unpolarized) leptonic tensor
%\begin{equation}
%L_{\mu\nu}^0=\frac{1}{2}{\rm Tr} (\hat k_2+m)\gamma_\mu
%(\hat k_1+m)\gamma_\nu
%\end{equation}
and for hadronic tensor
\begin{equation}
W_{\mu\nu}^u=\sum_i w_{\mu\nu}^i {\cal F}_i
\end{equation}
with
\begin{equation}
w_{\mu\nu}^1=-g_{\mu\nu}, \qquad
w_{\mu\nu}^2=\frac{p_{\mu}p_{\nu}}{M^2}
\end{equation}
and $\tau_p=Q^2/4M^2$,
\begin{equation}
{\cal F}_1=4 \tau_p M^2 G_M^2, \qquad
{\cal F}_2=4 M^2 \frac{G_E^2+\tau_p G_M^2}{1+\tau_p}
\end{equation}
It is convenient to define the contractions
\begin{eqnarray}
2\theta_B^1&=&L_{\mu\nu}^0 w_{\mu\nu}^1=2Q^2,
\\
2\theta_B^2&=&L_{\mu\nu}^0 w_{\mu\nu}^2=\frac{1}{M^2}
(S(S-Q^2)-M^2Q^2).
\end{eqnarray}

As a result for the Born cross section we have well known formula
\begin{equation}
\frac{d\sigma_0}{dQ^2}={2\pi\alpha^2 \over S^2 Q^4} \sum_i
\theta_B^i
{\cal F}_i,
\label{si0}\end{equation}
which is reduced to 
\begin{equation}
\frac{d\sigma_0}{dQ^2}\approx
{4\pi\alpha^2 \over Q^4}\frac{G_E^2+\tau_p G_M^2}{1+\tau_p}
\end{equation}
in ultrarelativistic approximation.

\subsection{Polarized part of cross section}
\label{polpart}

We consider two polarization measurements:
\begin{enumerate}
\item Initial proton is polarized and final electron is measured to
reconstruct $Q^2$. In this case there are four experimental situations
for asymmetry definition: target is polarized along (perpendicular) to
beam or $\vec q$ ($q=p_2-p$). Corresponding polarization 
4-vectors are denoted as
$\eta_L$ ($\eta_T$) or
$\eta_L^q$ ($\eta_T^q$).
\item Polarization and momentum of final proton are measured. Two
polarization states should be considered: final proton is polarized
along ($\eta'_L$) and perpendicular ($\eta'_T$) to  $\vec q$.
\end{enumerate}
If the polarization vector is kept in a general form the
polarization part of hadronic vector can be presented in the form
\begin{equation}
W_{\mu\nu}^p=\sum_{i=3}^4 w_{\mu\nu}^i {\cal F}_i
\end{equation}
with
\begin{equation}
w_{\mu\nu}^3=-i\epsilon_{\mu\nu\lambda\sigma}{q_{\lambda}\eta_{\sigma}
\over M}, \qquad
w_{\mu\nu}^4=i\epsilon_{\mu\nu\lambda\sigma}{q_{\lambda}p_{\sigma}\;
\eta q \over M^3}.
\end{equation}
For the consideration of the case of initial polarized particles
we have to choose the corresponding representation for the
polarization vector and structure functions in the
form
\begin{equation}
       {\cal{F}}_3=-2M^2G_EG_M, \qquad
       {\cal{F}}_4=-M^2G_M{G_E-G_M\over 1+\tau_p},
\label{sf34}\end{equation}
when the total hadronic tensor $W_{\mu\nu}=W_{\mu\nu}^u+W_{\mu\nu}^p$.
In the case of consideration of final polarization states
($\eta\rightarrow \eta'$) the same formulae for structure functions
(\ref{sf34}) but different sign for the last one (${\cal F}_4
\rightarrow -{\cal F}_4)$.

The 4--vector polarization $\eta$ can be expressed in terms of
4--momentum of particles in reaction.
The four considered cases correspond to four representations of
polarizations vectors
\begin{eqnarray}\label{polvec}
\eta_L&=&{1\over \sqrt{\lambda_s}}
\biggl(k_1-{S \over M} p\biggr),
\\
\eta_{T}&=& {1\over \sqrt{\lambda_s\lambda}}
\biggl((-SX+2M^2Q^2+4m^2M^2))k_{1} +
\nonumber \\ && \qquad
+ \lambda _{s}k_{2}
-(SQ^2+2m^{2}S_{x}) p\biggr)
\nonumber \\
\eta_{L}^q&=&{1\over \sqrt{\lambda_q}}
\biggl(2M(k_1-k_2)-{(S-X) \over M}p\biggr),
\nonumber \\
\eta^q_{T}&=&{1\over \sqrt{\lambda_q\lambda}}
\biggl( (2M^2Q^2-S_xX)k_1+
\nonumber \\ && \qquad
+(2M^2Q^2+S_xS)k_2
-Q^2(S+X)p\biggr),
\nonumber \end{eqnarray}
where
$\lambda =SXQ^2 - m^{2}\lambda_q-M^{2}Q^4$ and
$\lambda_q=S_x^2+4M^2Q^2$.
We note that the task of the calculation is reduced to contraction
of leptonic tensors at the born and RC level with
$w^{3,4}_{\mu\nu}$ using correspondent polarization vector
representation of general form
\begin{equation}\label{basis}
\eta=2(a_{\eta}k_1+b_{\eta}k_2+c_{\eta}p)
\end{equation}
and forthcoming integration. Also keeping in mind the case of RC we held
the variable $X$, which is $S-Q^2$ at the Born level and depends on
inelasticity $S-Q^2-v$ for RC; $S_x=S-X$. The definition of inelasticity
$v$ is given in the next section. It should be noted that we do
not consider
effects of normal polarization. It allows us to keep only three basis
vectors in Eq.(\ref{basis}).

Polarization part of the Born cross section is given by Eq.(\ref{si0}) for two
additional terms in the sum over $i$=3,4. Functions $\theta_i^B$ have the form
\begin{eqnarray}
\theta_3^B&=&{2m\over M}(q\eta\;k_2\xi-\eta\xi Q^2), \nonumber\\
\theta_4^B&=&{mQ^2q\eta\over M^3}(2p\xi-k_2\xi),
\end{eqnarray}
where lepton polarization vector can be taken as
\begin{equation}
\xi={2\over \sqrt{\lambda_s}}\bigl(\frac{S}{m}\; k_1-m\;p\bigr).
\end{equation}

If the $Q^2$ is calculated in terms of hadronic variables the
polarization vector  expansion looks like this
\begin{equation}
\eta'_{L,T}=2(a'_{L,T}k_1
+b'_{L,T}(p-p_2)+c'_{L,T}p)
\end{equation}
with
\begin{equation}
 a'_{L}=0, \;
 b'_{L}=-{Q^2+2M^2\over 2M\sqrt{\lambda_M}}, \;
 c'_{L}={Q^2\over 2M\sqrt{\lambda_M} }
\end{equation}
and
\begin{eqnarray}
 a'_{T}  & =&{Q^2(Q^2+4M^2)\over 2\sqrt\lambda_h \sqrt{\lambda_M}}, \\
 b'_{T}  &  =&{Q^2S+2M^2Q^2_u\over 2\sqrt\lambda_h  \sqrt{\lambda_M}},   \\
 c'_{T}  &  =&-{Q^2(2S-Q^2_u)\over 2\sqrt\lambda_h  \sqrt{\lambda_M}}.
\end{eqnarray}
Here
$\lambda_h
=SQ^2(S-Q^2_u)-M^2Q^4_u-m^2\lambda_M$
and
$
\lambda_M=Q^2(Q^2+4M^2)$. Similarly to the case above we hold the
variable $Q^2_u$, which is $Q^2$ and $Q^2+u$ for Born and RC cases
respectively.
The quantity $u$ is related with invariant mass of unobserved state. It
is also called inelasticity and is defined below (for the case of
hadronic variables see
(\ref{eq70})).

It easy to verify that 4--vector $\eta'_L$ and $ \eta'_T$
satisfy the necessary conditions:
$$ \eta'_Lp_2 = \eta'_Tp_2 =0, \ \ \eta'_L
\eta'_T=0,
   \ \ {\eta'_L}^2= {\eta'_T} ^2= -1. $$

In the rest frame of the scattered electron $p_2=(M,{\vec 0})$ the
vector of longitudinal polarization looks as
\begin{equation}
\eta'_L=(0,\vec n), \ \ \vec n^2=1 \ .
\end{equation}
The direction of 3--vector $\vec n$ coincides with the direction of
3--vector $\vec p_2$ in lab. system. Therefore, $\eta'_L$
indeed
describes the longitudinal polarization of the scattered proton. 4-vector
$\eta'_T$ has the form $(0,\vec m), \vec n\cdot\vec m =0$ in both
lab.system and rest frame system of the scattered proton. Thus, it
describes the transverse
 polarization in the scattered plane. It can
be defined up to sign only.

%Below we will concentrate on the calculation the spin--dependent part of
%the cross--section only. Another word, we will consider the ratio of the
%longitudinal polarization of the scattered electron to the
%perpendicular one.

In the case of longitudinal polarization contributes the term accompanied
with multiplier $G_M^2$ in the spin--dependent part of the cross-section,
because the part, that is proportional to $G_E G_M$ goes to zero for
$\eta=\eta^{\parallel}$. The situation is just contrary in the case
of transverse polarization. The simple calculation gives:
\begin{equation}\label{007}
\frac{\eta'_L}{\eta'_T} =
\frac{G_M}{G_E}\sqrt{\frac{-q^2}{M^2}}
\frac{k_1p+k_1p_2}{\sqrt{4(k_1p+k_1p_2)^2+q^2(4M^2-q^2)}}\ .
\end{equation}

It easy to verify that in the Breit system, where $p=(E,-\vec q/2),\
p_2 = (E,\vec q/2), \ q=(0,\vec q),$ the right side of Eq.(\ref{007}) coincides
(up to sign) with the expression given in \cite{Rekalo}. Indeed, in this system $\varepsilon_1 = \varepsilon_2 =\frac
{\sqrt{-q^2}}{2\sin{\theta_B/2}} \ (\theta_B$ is the electron scattering
angle in the Breit system) and
$$\frac{(k_1p+k_1p_2)}{\sqrt{4(k_1p+k_1p_2)^2+q^2(4M^2-q^2)}}=
   \frac{1}{2\cos{\theta_B/2}}\ . $$

\section{Radiative effects}

\subsection{Leptonic variables}

 For cross section of radiative process
\begin{equation} \label{radpro}
 e(k_1) + N (p) \longrightarrow e'(k_2) + \gamma(k) + N(p_2),
\end{equation}
we have again
\begin{equation}
d\sigma_r=\frac{M_r^2}{4pk_1}d\Gamma_r.
\end{equation}
The cross section of the process depends on $Q^2_l$ which for simplicity is
referred in this section as $Q^2$.
Phase space
\begin{equation}\label{phasespase}
d\Gamma_r=\frac{1}{(2\pi)^5}
\frac{d^3p_2}{2p_{20}}
\frac{d^3k_2}{2k_{20}}
\frac{d^3k}{2k_{0}} \delta(p+k_1-k_2-k-p_2)
\end{equation}
can be parameterized in terms of three variables: inelasticity
$v=\Lambda^2-M^2$ ($\Lambda=p+k_1-k_2$), 
$\tau=kq/kp$ and angle $\phi_k$ between planes
(${\bf q}$, ${\bf k}$) and  (${\bf k_1}$, ${\bf k_2}$).
Using the result (V.7.7) of \cite{Bukling_Kajanti} we have
\begin{equation}
d\Gamma_r={dQ^2\over 4(2\pi)^4S}
\int\limits_0^{v_m} \frac{dv}{4\sqrt{\lambda_q}}
\int\limits_{\tau_{min}}^{\tau_{max}} {d\tau}\frac{v}{(1+\tau)^2}
\int\limits_{0}^{2\pi} {d\phi_k},
\label{444}
\end{equation}
where (${\lambda_q=(v+Q^2)^2+4M^2Q^2}$) we use a variable $\tau$
instead of standard
$t$
\begin{equation}
t=Q^2+v-R; \qquad  R=\frac{v}{1+\tau}
\end{equation}
It allows to present final result in the form close to circle of papers
\cite{ASh,POLRAD20,AISh,Ak,AShS}.
Limits are defined as
\begin{eqnarray}
v_m&=&\frac{1}{2m^2}(\sqrt{\lambda_s}\sqrt{\lambda_m}-2m^2Q^2-Q^2S)=
\\
&=&
{2Q^2(\lambda_s-Q^2(S+m^2+M^2))\over Q^2S+2m^2Q^2+
\sqrt{\lambda_s}\sqrt{\lambda_m} }
\\ & \approx &
S-Q^2-{M^2Q^2\over S}
\end{eqnarray}
 and
\begin{equation}
2M^2\tau_{max,min}=v+Q^2\pm\sqrt{\lambda_q}
\end{equation}

Matrix element squared of radiated process is
\begin{equation}
M^2_r=\frac{e^6}{Q^4_h}L^r_{\mu\nu}W_{\mu\nu}
\end{equation}
The leptonic tensor of the radiative process is standard and 
can be found, for example, in \cite{AISh}. We note that here we use
more standard definition of the tensors (with additional factor 2
comparing to ones from the paper).
For contractions we have
\begin{equation}
L_{\mu\nu}^r w_{\mu\nu}^i=4\pi\sqrt{\lambda_q}
\sum_{j=1}^3 R^{j-3}\theta_{ij}.
\label{llrr}
\end{equation}
Functions $\theta_{ij}$ are similar to ones given in Appendix B of ref.
\cite{POLRAD20}. However there they are integrated over $\phi_k$. We keep this
integration because of possible dependence of acceptance function on this
angle.
Explicit form of the functions in our general case is discussed in Appendix.

We note that the well-known formula of soft photon approximation
is immediately obtained when keeping the term with $j=1$ and restricting
integration over (small photon energy) $v$ as $v_1<v<v_2\ll S$
(small photon energy):
\begin{equation}
{d\sigma_r \over dQ^2}={2\alpha \over
\pi}
(l_m-1)\log\frac{v_2}{v_1}
{d\sigma_0 \over dQ^2},
\end{equation}
where $l_m=\log(Q^2/m^2)$. 
For angular integration the formula (27)  of ref \cite{Ak} was
used.

Straightforward integration over photon phase space is not possible
because of infrared divergence. The first step of the treatment is the
identity transformation of the integrand
\begin{equation}
\sigma_R=\sigma_R-\sigma_{IR}+\sigma_{IR}
=\sigma_F+\sigma_{IR},
\label{ident}
\end{equation}
where $\sigma_F$ is finite for $k\rightarrow 0$ (here and below we use
the short notation for differential cross sections:
$\sigma_R\equiv {d\sigma_R}/{dQ^2}$ and so on). 
There is some
arbitrariness
in definition of $\sigma_{IR}$. Only asymptotical expression in the
limit  $k\rightarrow 0$ is fixed\footnote{There is once more limitation.
We must provide the conditions of applicability of the theorem about
changing of order integration and limitation. For example the uniform
convergence is required. Practically it means that we may subtract the
quantity with the same denominator}.
In our case we construct
$\sigma_{IR}$ using the term with $j=1$ in (\ref{llrr}) and formfactors
estimated in Born point. This term is factorized in front of Born cross
section as
\begin{equation}\label{FIR}
\sigma_0\frac{2}{\pi}\int \frac{d^3k}{2k_0}F_{IR},\quad
F_{IR}=\biggl( \frac{k_1}{2k_1k}-\frac{k_2}{2k_2k} \biggr)^2.
\end{equation}
As a result infrared part can be written in the factorized form
\begin{equation}
\sigma_{IR}={\alpha \over\pi}\delta^{IR}_{R}\sigma_0
={\alpha \over\pi}(\delta_{S}+\delta_{H})\sigma_0.
\end{equation}
The quantities $\delta_{S}$ and $\delta_{H}$ appear after additional
splitting the integration region over inelasticity $v$ by the
infinitesimal parameter $\bar v$
\begin{eqnarray}
\delta_{S}&=&\frac{-1}{\pi}\int\limits_0^{\bar
v}dv\int\frac{d^{n-1}k}{(2\pi\mu)^{n-4}k_0}F_{IR}\delta((\Lambda-k)^2-M^2),
\nonumber \\
\delta_{H}&=&\frac{-1}{\pi}\int\limits_{\bar
v}^{v_m}dv\int\frac{d^3k}{k_0}F_{IR}\delta((\Lambda-k)^2-M^2).
\end{eqnarray}
We note that in contrast to Mo and Tsai formalism this artificial
parameter $\bar v$ completely cancels in final expressions.
The way to calculate these integral was offered in \cite{BSh} (see also
\cite{Ak} and review \cite{Bardin}). In our case we have
\begin{eqnarray}
\delta_S&=&2\biggl(P_{IR}+\log\frac{\bar v}{\mu M}\biggr)(l_m-1)\!+\!
\log\frac{S(S-Q^2)}{m^2M^2}+S_{\phi},
\nonumber \\
\delta_H&=&2(l_m-1)\log\frac{v_m}{\bar v}.
\end{eqnarray}
These contributions have to be added with
result for vertex function which is standard
\begin{eqnarray}\label{deltav}
\delta_V&=&-2\biggl(P_{IR}+\log\frac{m}{\mu}\biggr)(l_m-1)
\\ && \qquad
-\frac{1}{2}l_m^2
+\frac{3}{2}l_m-2+\frac{\pi^2}{6}. \nonumber
\end{eqnarray}
For this sum we have
\begin{equation}
\frac{\alpha}{\pi}\biggl( \delta_S+\delta_H+\delta_V \biggr)
=\delta_{inf}+\delta_{VR},
\end{equation}
where
\begin{eqnarray}
\delta_{inf}&=&{\alpha\over \pi}\bigl( l_m-1 \bigr) \log \frac{v_m^2}{S(S-Q^2)},
\\
\delta_{VR}&=&{\alpha\over\pi}\biggl(\frac{3}{2}l_m-2-\frac{1}{2}\log^2\frac{S}{S-Q^2}
\nonumber \\ &&
+{\rm Li}_2\biggl( 1-\frac{M^2Q^2}{S(S-Q^2)}\biggr )-\frac{\pi^2}{6}\biggr).
\nonumber
\end{eqnarray}
Here we used ultrarelativistic expression for function $S_{\phi}$ from
\cite{Sh}.

Finally the cross section that	takes into account radiative effects can
be written as
\begin {equation}
\sigma _{obs} = \sigma _0 e^{\delta_{inf}}
(1+ \delta_{VR}+\delta_{vac})+\sigma_{F}.
\label{eq1}
\end {equation}
Here the corrections
 $\delta_{inf}$  and   $\delta_{vac}$  come from  radiation of soft
photons and effects of vacuum polarization.
The correction	$\delta_{VR}$ is an infrared-free sum of factorized
parts of real and virtual photon radiation. $\sigma_F$ is the 
infrared-free contribution of bremsstrahlung
process:

\begin{eqnarray}
\label{sir}
\sigma_F&=&-{\alpha^3 \over 2S^2}
\int\limits_0^{v_m} {dv}
\int\limits_{\tau_{min}}^{\tau_{max}}
\frac{d\tau}{1+\tau}
\int\limits_{0}^{2\pi} {d\phi_k} \times
\\ && \qquad\times
\sum_i
\biggl[\sum_{j=1}^3 {\cal A} R^{j-2}\theta_{ij} {{\cal F}_i \over Q^4_h}
 -4F^0_{IR}\theta_{i}^B {{\cal F}_i^0 \over RQ^4_l}\biggr]\nonumber
.
\end{eqnarray}
Here $\cal A$ is integrated over the $\phi$ acceptance function.

\subsection{Hadronic variables}

 For the cross section of radiative process
\begin{equation}
 e(k_1) + N (p) \longrightarrow e'(k_2) + \gamma(k) + N(p_2),
\end{equation}
we have again
\begin{equation}
d\sigma_r=\frac{M_r^2}{4pk_1}d\Gamma_r
\end{equation}

The parameterization of photonic phase space, integration over it
developed for
so-called hadronic emission within considered Bardin and Shumeiko
approach can be applied to this case, if we picture the
corresponding Feynman graph upside down. Thus phase space (\ref{phasespase})
%\begin{equation}
%d\Gamma_r=\frac{1}{(2\pi)^5}
%\frac{d^3p_2}{2p_{20}}
%\frac{d^3k_2}{2k_{20}}
%\frac{d^3k}{2k_{0}} \delta(p+k_1-k_2-k-p_2)
%\end{equation}
can be parameterized in terms of three invariant variables
\cite{BarKal,AISh2}: inelasticity
$u=(k_2+k)^2-m^2=2k_2k$, $w=2k_1k$ and $z=2p_2k$
\begin{equation}\label{eq70}
d\Gamma_r={dQ^2 d\phi\over (4\pi)^4S}
\int\limits_0^{u_m} {du}
\int\limits_{w_{min}}^{w_{max}} {dw}
\int\limits_{z_{min}}^{z_{max}} \frac{dz}{\pi\sqrt{-R_z}},
\label{555}
\end{equation}
where $R_z$ comes from Gram determinant ($16R_z=\Delta(k_1,p,p_2,k)$)
and coincides
with standard $R_z$-function appearing in Bardin-Shumeiko approach
\cite{BSh,BarKal}. Explicitly
\begin{equation}
R_z=A_zz^2-2B_zz+C_z.
\end{equation}
For completeness we give coefficients in our 
notation\footnote{$Q^2$ in this subsection is $Q^2_h$}
\begin{eqnarray}
A_z&=&\lambda_q=      (u+Q^2)^2+4Q^2m^2, \\
B_z&=&u(u+Q^2)s_q-(u-Q^2)Sw-2m^2Q^2(u-w), \nonumber \\
C_z&=& (us_q-Sw)^2+4M^2uw(Q^2 + u - w)
\nonumber \\ &&
-4M^2m^2(u-w)^2, \nonumber
\end{eqnarray}
where $s_q=S-Q^2+w-u$.
%(${\lambda_q=(v+Q^2)^2+4*M^2Q^2}$)

We note that we introduced invariant variable $z$ which corresponds to
azimuthal angle in (\ref{444}). It is more convenient for introducing
explicit expressions for experimental cuts.

Limits are defined as
\begin{eqnarray}
u_m&=&\frac{1}{2M^2}(\sqrt{\lambda_s}\sqrt{\lambda_M}-2M^2Q^2-Q^2S)=
\\ &=&
{2Q^2(\lambda_s-Q^2 (S+m^2+M^2))\over Q^2S+2M^2Q^2+
\sqrt{\lambda_s}\sqrt{\lambda_M} }
\end{eqnarray}
 and
\begin{equation}\label{limw}
w_{max,min}=\frac{u}{2(u+m^2)}(Q^2+u+2m^2\pm\sqrt{\lambda_q}).
\end{equation}
Limits $z_{min,max}$ are defined as solutions of the equation $R_z=0$.
\begin{equation}
\lambda_q z_{max,min}=B_z\pm\sqrt{D},
\end{equation}
where
\begin{eqnarray}
D &=& 4(M^2\lambda_q + m^2Q^4 - Q^2S(S-Q^2-u))\times
 \nonumber \\ &&
\times(m^2(w-u)^2 + uw(w-u-Q^2)).
\end{eqnarray}
 Two solutions of the equation $D=0$ give limits on $w$ (\ref{limw}).

Matrix element squared of radiated process is calculated as
\begin{equation}
M^2_r=-\frac{e^6}{Q^4}L^r_{\mu\nu}W_{\mu\nu}=
-\frac{e^6}{Q^4}(T_{IR}+T_3+T_4+T_{34})
\end{equation}
where
\begin{eqnarray}
T_3&=&{4a'_{L,T}{\cal F}_3\over M S}
\biggl(2{m^2\over w^2}(Q^4z-2Q^2Su+Q^2uz-u^2S)
\nonumber\\&&+{SQ^2\over u}(w-2Q^2)
+{S\over w}(4Q^4+3Q^2u+u^2)
%\nonumber \\&&
-Sw\biggr),
\nonumber \\
T_4&=&{-2Q^2a'_{L,T}{\cal F}_4\over M^3 S}
\biggl({2m^2\over w^2}
(Q^2Su-Q^2uz-2S^2u
+Su^2
\nonumber \\&&
+2Suz-2uz^2)
+{S\over w}
(-2Q^4+4Q^2S-2Q^2u
\nonumber \\&&
-2Q^2z+2Su-u^2)
 -2S^2+Sw+2Sz\biggr),
\nonumber \\
T_{34}&=&{2Q^2\over M^3S}
(a'_{L,T}{\cal F}_4Q^2-2M^2c'_{L,T}{\cal F}_3-
\nonumber \\&& \qquad
-2b'_{L,T}{\cal F}_4Q^2-c'_{L,T}{\cal F}_4Q^2)\times
\nonumber \\&&
\times\biggl(2{m^2\over w^2}
(Q^2z-Su-2Sz+2z^2)
+{2SzQ^2\over uw}
\nonumber \\&&
+{S\over u}(2S-w-2z)
+{S\over w}
(2Q^2-2S+ u)
 \biggr).
\nonumber
\end{eqnarray}

The contribution $T_{IR}$ is
\begin{eqnarray}
T_{IR}&=&4\bigl({m^2\over w^2}+{m^2\over u^2}-{Q^2\over uw}\bigr)T_0,
\nonumber \\
T_0&=&-{Q^2\over M^3}
 \bigl(  2M^2{\cal F}_3(a'_{L,T}Q^2-c'_{L,T}Q^2+2c'_{L,T}S)
\nonumber \\ &&
+ Q^2(Q^2-2S){\cal F}_4(a'_{L,T}-2b'_{L,T}-c'_{L,T})   \bigr).
\end{eqnarray}
Structure functions ${\cal F}_3$ and ${\cal F}_4$ are defined in (\ref{sf34})
and are functions of $Q^2_h$.
We note that Born cross section can be written in terms of $T_0$ for $a_\eta$,
$b_\eta$, $c_\eta$ taken for $u\rightarrow 0$
\begin{equation}
      \sigma_0={\pi\alpha^2 \over Q^4S^2}T_0^B.
\end{equation}

The radiative cross section can be given as
\begin{equation}
\sigma_R={-\alpha^3 \over 4S^2Q^4} \int {dudwdz \over \pi\sqrt{-R_z}}
{\cal A} (T_{IR}+T_3+T_4+T_{34}).
\end{equation}
Performing the explicit integration over the region of small energies
($0<u_1<u_2\ll m^2,M^2,Q^2,S$) we obtain the well-known result in the soft 
photon limit:
\begin{eqnarray}
\sigma_R&=&-{\alpha\over \pi} \sigma_0 \int\limits_{u_1}^{u_2}du
\biggl[
\frac{1}{u}+{m^2\over u(u+m^2)}-{Q^2\over u\sqrt{\lambda_q}} 
\log{w_{max}\over w_{min}} \biggr] 
\nonumber
\\ &=& {2\alpha\over \pi} 
\log{u_2\over u_1} (l_m-1)
\sigma_0.
\end{eqnarray}

The radiative cross section has infrared divergence, so
for this case the identity transformation like (\ref{ident}) has to be
performed also. However formfactor is not dependent on photon variables,
so only the acceptance function should be subtracted. Corresponding
integrals looks similar.
As a result infrared part can be written in the factorized form
\begin{equation}
\sigma_{IR}={\alpha \over\pi}\delta^{h,IR}_{R}\sigma_0
={\alpha \over\pi}(\delta_{S}^h+\delta_{H}^h)\sigma_0.
\end{equation}
The quantities $\delta_{S}$ and $\delta_{H}$ appear after additional
splitting the integration region over inelasticity $v$ by the
infinitesimal parameter $\bar v$
\begin{eqnarray}
\delta_{S}^h&=&\frac{-1}{\pi}\int\limits_0^{\bar
u}du\int\frac{d^{n-1}k}{(2\pi\mu)^{n-4}k_0}F_{IR}\delta((\Lambda_h-k)^2-m^2),
\nonumber\\
\delta_{H}^h&=&\frac{-1}{\pi}\int\limits_{\bar
u}^{u_m}du\int\frac{d^3k}{k_0}F_{IR}\delta((\Lambda_h-k)^2-m^2),
\end{eqnarray}
where $\Lambda_h=k_1+p-p_2$ and $F_{IR}^h$ is defined in eq.(\ref{FIR}).
The integration gives the following explicit results
\begin{eqnarray}
\delta_S^h&=&2\biggl(P^{IR}+\log\frac{\bar
u}{m\mu}\biggr)(l_m-1)+1+l_m-l_m^2
-\frac{\pi^2}{6},
\nonumber\\
\delta_H^h&=&2(l_m-1)\log\frac{u_m}{\bar u}-{1\over 2}
\log^2{u_m \over m^2}
+\log{u_m \over m^2}
\nonumber\\&&
-l_w(l_m+l_w-l_v)
-\frac{\pi^2}{6}-{\rm Li}_2\biggl(-{u_m
\over Q^2}\biggr),
\nonumber\\&&
l_v=\log\frac{u_m}{Q^2}, \qquad
l_w=\log\biggl(1+\frac{u_m}{Q^2}\biggr).
\end{eqnarray}
Sum of them and the contribution of vertex function gives again
infrared free result
\begin{eqnarray}
\delta_{VR}^h&=&l_m(l_v-l_w+\frac{3}{2})-l_v-1-\frac{3}{2}l_w^2-\frac{1}{2}l_v^2+2l_wl_v
\nonumber\\&&
      -{\rm Li}_2\biggl({Q^2\over Q^2+u_m}\biggr).
\end{eqnarray}

The cross section that	takes into account radiative effects can
be written as
\begin {equation}
\sigma _{obs} = \sigma _0 %e^{\delta_{inf}}
(1+ \delta_{VR}^h+\delta_{vac})+\sigma_{F}^h.
\label{eq1b}
\end {equation}

The explicit expression for $\sigma_{F}^h$ is
\begin{eqnarray}
\sigma_{F}^h&=&-{\alpha^3\over 4Q^4S^2}
\int\limits_0^{u_m} {du}
\int\limits_{w_{min}}^{w_{max}} {dw}
\int\limits_{z_{min}}^{z_{max}} \frac{dz}{\pi\sqrt{-R_z}}
\\ && \qquad
\biggl[{\cal A}(T_{IR}+T_3+T_4+T_{34}) - T_{IR}^B \biggr].
\end{eqnarray}
with $T_{IR}^B =T_{IR}^B(T_0\rightarrow T_0^B)$

\subsection{Kinematical cuts}\label{kincuts}

In this section we show how experimental cuts can be introduced to this
approach. As an example we consider conditions of experiment
\cite{1CEBAFothers} at
CEBAF.

\begin{figure}[!ht]
\unitlength 1mm
\begin{picture}(80,60)
\put(0,5){
\epsfxsize=6.4cm
\epsfysize=6.4cm
\epsfbox{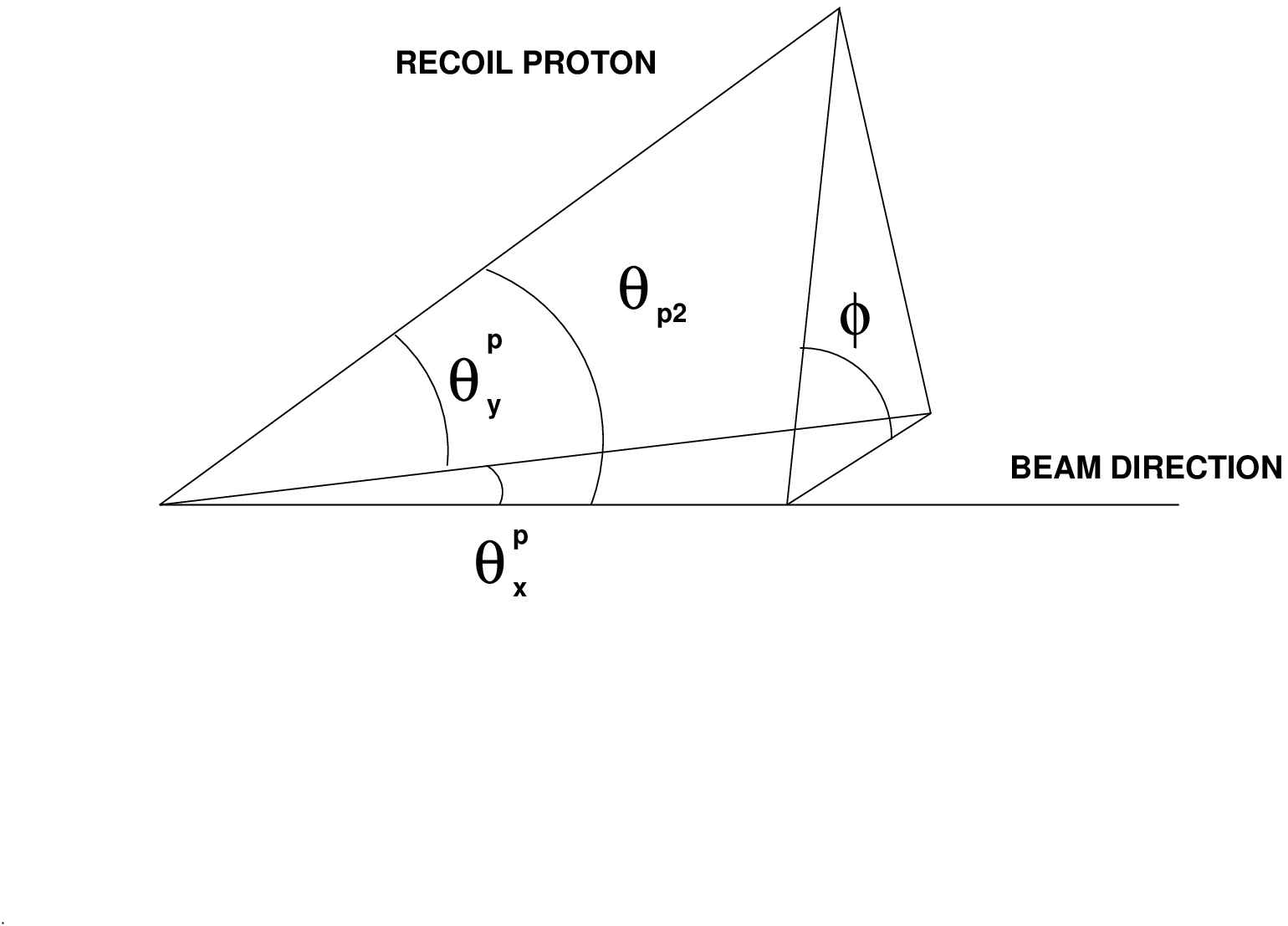}
}
\end{picture}
\vspace{-0.5cm}
\caption{Recoil proton angles
definitions. The beam and the hadron-arm detector (HRS) define the
horizontal plane.
}
\label{Fig1}
\end{figure}

The following restriction on momentum of final electron and proton have
to be done. We consider high-resolution spectrometer (HRS)
\cite{1CEBAFothers} as a rectangular area with some energy acceptance. 
We describe this rectangle by two angles between corresponding
planes: $\theta_x^{e,p}$ and $\theta_y^{e,p}$. For one of the scattered particle
these angular definitions are given in Figure \ref{Fig1}. The 
upper index corresponds
to detected particle: electron or proton. The angles and momenta of
final particles being measured in lab.frame have to be expressed in
terms of kinematical invariants. The simplest way to do it is to apply a
formalism of Gramm determinants, which detailed description can be found
in Ref.\cite{Bukling_Kajanti}. We will give starting expressions in terms
of
four-momenta and Gramm determinants as well as explicit results for the
used (in (\ref{555})) invariant variables. In terms of Gramm
determinants the momenta (in the rest frame of $p$) of
final proton and electron are given by the formula
       \begin{equation}
 |{\bf p_2}|^2=-{\Delta(p,p_2)\over p^2}, \quad
 |{\bf k_2}|^2=-{\Delta(p,k_2)\over p^2}.
\end{equation}
It gives immediately for the Born process
\begin{equation}
 |{\bf p_2}|={\sqrt{\lambda_q}\over 2M}, \quad
 |{\bf k_2}|={\sqrt{\lambda_z}\over 2M}
\end{equation}
both for Born and radiative process. Here $\lambda_z=
(S-Q^2_h-z_1)^2-4M^2m^2$.

Cosine of polar angle of the direction of final proton (with respect to
the
beam direction) is defined as
\begin{equation}
 \cos\theta_{p2}={k_1p\;p_2p-p^2\;p_2k_1\over
\Delta(p,k_1)\Delta(p,p_2) }
\end{equation}
It gives
\begin{equation}
\sin^2\theta_{p2}={4\lambda_{sx}M^2\over \lambda_M S^2}.
\end{equation}
where $\lambda_{sx}=SXQ^2-S_x^2M^2$.
We have to use
$X=S-Q^2$ and $S_x=Q^2$ for Born process
$X=S-Q^2-u$ and $S_x=Q^2+u$ for radiative one.
In terms of the angle the horizontal and vertical angles of proton momentum are
\begin{eqnarray}
\sin\theta_y^p&=&\sin\phi\;\sin\theta_{p2},
\nonumber \\
\tan\theta_x^p&=&\cos\phi\;\tan\theta_{p2},
\end{eqnarray}
where $\phi$ was introduced in (\ref{phiall}).
%We note that $\theta_y^p=\phi$ for all cases.
%The definition of the horizontal angle of final electron will complete
%the cuts definition at the born level.
%It can be obtained by the similar way
%\begin{equation}
%\sin^2\theta_{k2}={4Q^2_e\lambda_{qz}M^2\over
%S_{qz}^2S^2}.
%\end{equation}
%Again the horizontal angle is defined as
%\begin{equation}
%\sin\theta_x^e=\cos\phi\;\sin\theta_{k2}
%\end{equation}

At the Born level all vectors ({$\bf p_2$, $\bf k_1$ and $\bf k_2$}) are
in the same plane. However for the unobserved photon there is a nonzero
angle $\Delta\phi$ between planes ($\bf k_1,p_2$) and ($\bf k_1,k_2$).
In terms of Gramm determinants it is defined as

\begin{equation}
 \cos\Delta\phi={G\left(
\begin{array}{ccc}
 \displaystyle
 p & k_1 & p_2 \\
\displaystyle
 p & k_1 & k_2
 \end{array}
\right)
\over
\Delta(p,k_1,k_2)\Delta(p,k_1,p_2) }
\end{equation}
Explicitly we have
\begin{eqnarray}
\sin^2\Delta\phi&=&
-{S^2
\over
4Q^2_e\lambda_{sx}\lambda_{qz}}
\bigl[ (S(u - w)-z_1u+Q^2w)^2
\nonumber\\&&
\!\!+4Q^2_eM^2uw
+Q^2(2w+z_1)z_1(Q^2+u)
\nonumber\\&&
\!\!+z_1^2uQ^2+2z_1Q^2uw
-2Sz_1(u+w)Q^2\bigr]
\end{eqnarray}
where
%\begin{equation}
$\lambda_{qz}=SS_{qz}-2M^2Q^2_e$, $S_{qz}=S-Q^2-z_1$, $Q_e^2=Q^2+u-w$ and
$z_1=z+u-w$.
%\end{equation}
Now we can define angles of final electron for radiative process
\begin{eqnarray}
\sin\theta_y^e&=&\sin(\phi+\Delta\phi)\;\sin\theta_{k2},
\nonumber \\
\tan\theta_x^e&=&\cos(\phi+\Delta\phi)\;\tan\theta_{k2},
\end{eqnarray}
%
%
%
%\begin{eqnarray}
%\theta_y^e&=&\phi+\Delta\phi\\
%\sin\theta_x^e&=&\cos(\phi+\Delta\phi)\sin\theta_{k2}
%\end{eqnarray}
where $\theta_{k2}$ is polar angle of final electron. For it we have
\begin{equation}
\sin^2\theta_{k2}={4Q^2_e\lambda_{qz}M^2\over S_{qz}^2S^2}.
\end{equation}

\section{Numerical results}

In this section we present the FORTRAN code MASCARAD (Subsection \ref{MASCARAD})
developed on the basis of
formalism presented in the last sections. This code uses Monte-Carlo methods to
calculate radiative correction to the observable quantities in polarized ep
scattering measurements. The numerical results of applying this code
for kinematical conditions of Jlab are given in Subsections \ref{Numlep} and
\ref{Numhad} with leptonic and hadronic variables respectively. In
last Subsection we discuss the influence of
experimental cuts on observables quantities in polarized scattering.
% and relationship between observables in asymmetry and recoil
%polarization measurements.

\subsection{FORTRAN code MASCARAD}\label{MASCARAD}

There are two variants of the code: {\bf mascarad\_l.f} and {\bf mascarad\_h.f}
dealing with leptonic and hadronic variables respectively. First code does not
require any external libraries. However the histogramming by HBOOK can be
optionally included in
second
variant. In this case {\bf mascarad\_h.f} 
requires CERNLIB installed. In external file one
can choose kinematical variables, accuracy of calculation and value of a cut on
inelasticity. An option to include kinematical cuts described in Section
\ref{kincuts} is available for {\bf mascarad\_h.f}.
As an output one has
the value of the Born cross section, radiative correction factor
(with estimation
of statistical error) in chosen kinematical points.

\subsection{Numerical results. Leptonic variables}\label{Numlep}

Absolute and 
relative
corrections to asymmetry can be defined as (see (\ref{first}))
\begin{eqnarray}\label{deldef}
\Delta A_i&=&A_i-A_{i0}={(1+\delta)\sigma_0^p+\sigma_R^p \over
(1+\delta)\sigma_0^u+\sigma_R^u} - {\sigma_0^p  \over
\sigma_0^u }\\
\Delta_i&=&{A_i-A_{i0} \over A_{i0}}=
{\delta_p-\delta_u \over 1+\delta+\delta_u}
\end{eqnarray} 
where index $i$ runs over all considered cases: $i=L,T,qL,qT$;
$\delta_{u,p}=\sigma_{R}^{u,p}/\sigma_{0}^{u,p}$.  Here the correction $\delta$
is usually large because of contributions of leading logarithms. However it 
exactly canceled in numerator of expression for correction to asymmetry. That
is a reason why the correction to cross section can be large, and correction to
asymmetry is relatively small.

\begin{figure}[!b]
\unitlength 1mm
\begin{picture}(80,90)
\put(-7,5){
\epsfxsize=9.6cm
\epsfysize=9.6cm
\epsfbox{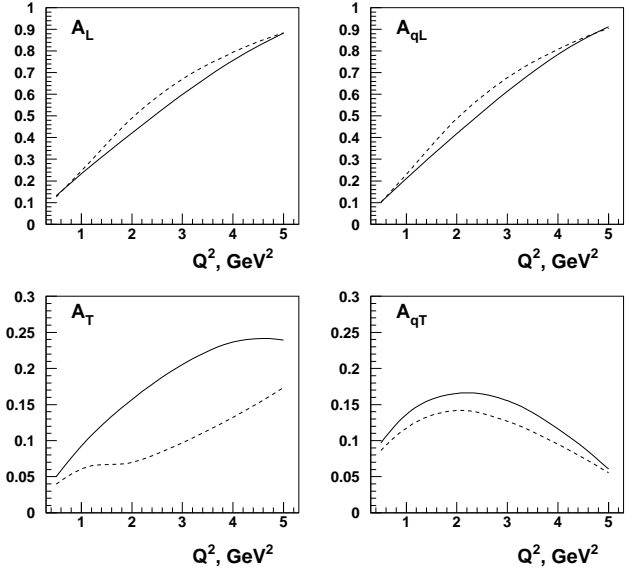}
}
\end{picture}
\vspace{-1cm}
\caption{Born (solid line) and observed (dashed line) asymmetries vs
$Q^2$. No kinematical cuts on inelasticity ware used. $E=4$ GeV.}
\label{Fig2}
\end{figure}

Born and observed asymmetries are presented in Figure \ref{Fig2}. Four
lines corresponds to four considered cases defined in sections
\ref{polpart}. No cuts were used for the missing mass. As a
result hard photon contributions gives different contributions to
spin averaged and spin dependent part of cross sections due to
unfactorizing properties. For longitudinal asymmetries $\delta_p>\delta_u$ and
there are positive contributions to RC. Opposite situation with transverse
asymmetries.

One can see that the transverse asymmetry $A_T$ in respect to beam direction 
has large
correction. It is not in contradiction with other plots in this figure. 
The polarized
part of cross sections in the cases (L,T)
and (qL, qT) are related to
each other by some
unitary transformation
\begin{equation}\label{cosgamma}
{d\sigma^p_{t}\over dQ^2dv}=
\cos\gamma{d\sigma^p_{qt}\over dQ^2dv} 
-\sin\gamma{d\sigma^p_{ql}\over dQ^2dv}.
\end{equation} 
Because of dependence of polarization vectors on inelasticity 
(see (\ref{polvec}) and definitions there) 
this angle $\gamma$ is a function of $v$
\begin{equation}
\sin^2\gamma={4M^2\lambda\over \lambda_s\lambda_q}
\quad
\cos \gamma={SS_x+2M^2Q^2\over \sqrt{\lambda_s\lambda_q}}
\end{equation}
and only
unintegrated cross sections are related as ({\ref{cosgamma}).  This sine hardly
suppresses the cross section of the hard photon. Weighed with the sine and
cosine, cross sections in r.h.s of (\ref{cosgamma}) have the same
signs, similar magnitude and compensate each other. As a result 
$\delta_p \ll \delta_u $  and the asymmetry $A_{T}$ has a large negative
contribution.
 
In practice RC to the asymmetries can be
essentially reduced by applying the cut on missing mass or inelasticity
which is also measured quantity in elastic electron-proton scattering.
In Figure \ref{Fig3} it is shown how these relative corrections depend on
value of the cut on missing mass or inelasticity.

\begin{figure}[!ltb]
\unitlength 1mm
\begin{picture}(80,80)
\put(0,5){
\epsfxsize=8cm
\epsfysize=8cm
\epsfbox{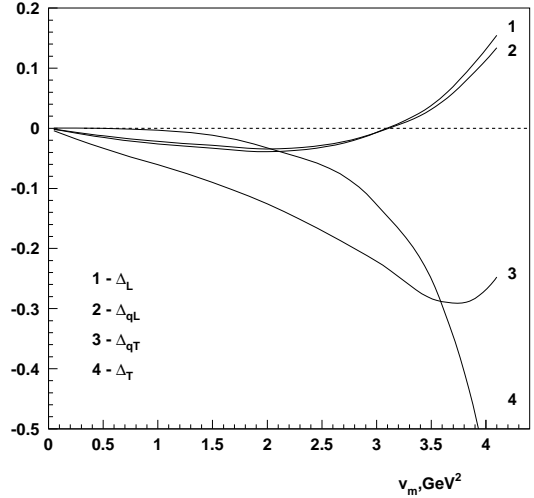}
}
\end{picture}
\caption{Relative RC to asymmetries defined in (\ref{deldef}) vs value of
inelasticity cut for $Q^2$=3 GeV$^2$; $E=4$ GeV.}
\label{Fig3}
\end{figure}

\subsection{Numerical results. Hadronic variables}\label{Numhad}

Similarly to the case of leptonic variables let us define relative RC
to the ratio of recoil nucleon polarizations $P_T/P_L$ as
\begin{equation}
\Delta={P_T/P_L-P_T^0/P_L^0\over P_T^0/P_L^0}
\end{equation}
In Figure \ref{Fig4} this correction is given for several values of cut on
missing mass. It can be seen that radiative correction in this case is
smaller then in the case of leptonic variables.

However in practice experimental situation can be complicated then
applying single kinematical cut on missing mass. In the last
and future experiments
\cite{1CEBAFothers,2CEBAFothers} at JLab on measurement ratio of elastic
formfactors of proton the all events appearing in
detectors\footnote{Here we do not consider possible losing events in
detector. By
the other words we take into account only geometrical acceptance. However
influence of apparatus acceptance also can be taken into account in our
approach by introducing some map of device acceptance to MASCARAD. There
is a certain place in the code to do it.} were accepted
for analysis.  To calculate RC in this situation we have to apply all cuts
discussed in previous section. The Table \ref{Tab1} gives the results for
the
last \cite{1CEBAFothers} and future \cite{2CEBAFothers}  
experiments (upper and below the intersection line).

As we can see the RC does not exceed 1\%.

\begin{figure}[!ltb]
\unitlength 1mm
\begin{picture}(80,80)
\put(0,5){
\epsfxsize=8cm
\epsfysize=8cm
\epsfbox{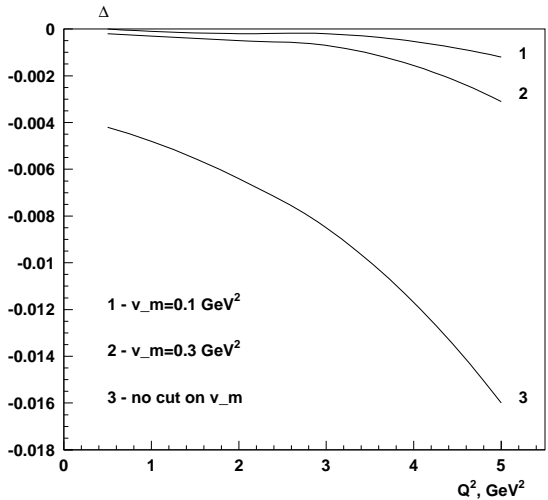}
}
\end{picture}
\caption{Relative RC to the ratio of recoil proton polarizations 
vs $Q^2$ for three
values of inelasticity cut.}
\label{Fig4}
\end{figure}

\vspace{2cm}

\begin{table}[!ht]
\caption{The results for RC to asymmetry ${P_T/P_L}$}
\label{Tab1}
\begin{tabular}{ccc}
%\hline
 $E$,GeV & $Q^2$,GeV$^2$ & $\Delta$ \%	  \\
\hline
0.934	 &   0.45--0.53  & -1.01	\\
0.934	 &   0.77--0.81  & -1.54	\\
1.821	 &   1.11--1.25  & -0.95	\\
3.395	 &   1.37--1.59  & -0.59	\\
3.395	 &   1.65--1.89  & -0.64	\\
4.087	 &   1.75--2.01  & -0.62	\\
4.090	 &   2.30--2.64  & -0.71	\\
4.087	 &   2.77--3.17  & -0.80	\\
4.090	 &   3.27--3.67  & -0.95	\\
\hline
4.845	 &  3.5 	 &  -0.80     \\
4.845	 &  4.2 	 &  -0.96     \\
5.545	 &  4.9 	 &  -0.95     \\
6.045	 &  5.6 	 &  -0.97     
%\hline
\end{tabular}
\end{table}

%\vspace{1cm}

\section{Discussion and Conclusion}

In this paper we applied the approach of Bardin and Shumeiko for calculation 
of the model independent radiative correction of
the lowest order in processes of elastic electron-proton scattering. Current
experiments on the process measure different polarization observables such as
spin asymmetries in different configurations of  polarizations of initial
particles and the ratio of recoil proton polarizations allowing to access the
ratio of electromagnetic formfactors of the proton. That is way special
attention was paid to radiative correction to polarized parts of the cross
section. 

The chosen method of calculation allowed to obtain explicit formulae
in the cases of so-called electron and hadron variables. They correspond to the
cases when kinematics of the measured process is reconstructed from the momentum
of final electron and proton respectively. It was shown that in spite of the
formulae for Born case are exactly the same for both cases, 
all ingredients of the
RC calculation are different in these cases. Physical (or kinematical) reason of
it is the fact that 
in the first case the radiating particle (electron) is measured, but in the
second case it escapes unmeasured or is integrated over.

Explicit formulae for the lowest order radiative correction are exact if not to account
ultrarelativistic approximation taking  off the terms $\sim m^2/Q^2$. However it
is not a limitation of the approach. These terms can be easily restored in the
case of need (for example, if they will apply to to muon scattering and the
accuracy of $\sim m_{\mu}^2/Q^2$ will be required.    

Contrary to the case of inclusive DIS the integration here is left for numerical
analysis. All integrals are finite after used the procedure of covariant
cancellation of infrared divergences. This form of answer allows to include  
acceptance effects
to the
integrand. The function of acceptance usually depends on
final angles and momenta, which can be expressed in terms of integration
variables. Corresponding way to do it is discussed in Section \ref{kincuts}.

%\penalty -1000000

%\vspace{-3cm}

On the basis of the exact formulae the FORTRAN package MASCARAD was developed.
It includes codes both for the electron and hadron variable measurement.
The applying of the package to the radiative correction 
procedure allows to provide the account of the model independent correction 
in current measurements (including polarization) of elastic electron-nucleon
scattering. Our numerical analysis shows that radiative effects can be
important especially in the cases of transversely polarized targets. However
using  kinematical cuts such as single cut on inelasticity or a cut on 
kinematical
variables of second (undetected) particle allows to reduce the effect
essentially.

\section*{Acknowledgements}
We would like to acknowledge useful discussion with A.Ilyichev.
We thank our colleagues at Jefferson Lab for useful discussions.
We thank the US Department of Energy for support under contract
DE-AC05-84ER40150. Work of NM was in addition supported by Rutgers
University
through NSF grant PHY 9803860 and Ukrainian DFFD.

\section*{Appendix}
\label{sec:appena}
\setcounter{equation}{0}
\renewcommand{\theequation}{A.\arabic{equation}}

The functions $\theta_{ij}$ are defined the same way as in Appendix B of
ref.\cite{ASh} or with more details in in Appendix B of
ref.\cite{POLRAD20}. However there the formulae are integrated over
photon azimuthal angle (or equivalently over $z$, see (B7) of
ref.\cite{ASh}). Below we define the procedure how the explicit form of
this functions can be written in out case.

 All formulae (B1-B2) of ref.\cite{ASh} or (B.1-B.11)
 of ref.\cite{POLRAD20} can be kept for our case. Instead of functions
$F's$ from (B5) \cite{ASh} or (B.12) \cite{POLRAD20} we use the
following expressions

\begin{equation}
F_{d}=\frac{F}{z_1z_2},
F_{1+}=\frac{F}{z_1}+\frac{F}{z_2},
F_{2\pm}=F\biggl(\frac{m^2}{z_2^2}\pm\frac{m^2}{z_1^2}\biggr),
\end{equation}
where $F=1/(2\pi\sqrt{\lambda_Q})$ and $F_{IR}=F_{2+}-Q^2F_d$.

\begin{eqnarray}
z_1&=&\frac{1}{\sqrt{\lambda_q}}(Q^2S_p+\tau(SS_x+2M^2Q^2)-2M\sqrt{\lambda_z}\cos\phi_{k})
\nonumber \\
z_2&=&\frac{1}{\sqrt{\lambda_q}}(Q^2S_p+\tau(XS_x-2M^2Q^2)-2M\sqrt{\lambda_z}\cos\phi_{k})
\nonumber
\end{eqnarray}
where
\begin{equation}
\lambda_z
       =(\tau-\tau_{min})(\tau_{max}-\tau)(SXQ^2-M^2Q^4-m^2\lambda_q)
\end{equation}

\end{document}